\documentstyle[aps,prl,psfig]{revtex}
\newcommand{\dd}{{\rm d}}
\newcommand{\erfc}{\text{erfc}}
\newcommand{\avg}[1]{\langle{#1}\rangle}
\newcommand{\cro}[1]{\left[{#1}\right]}
\newcommand{\e}{\text{e}}

\newcommand{\be}{\begin{equation}}
\newcommand{\ee}{\end{equation}}
\newcommand{\beas}{\begin{eqnarray*}}
\newcommand{\eeas}{\end{eqnarray*}}
\newcommand{\bea}{\begin{eqnarray}}
\newcommand{\eea}{\end{eqnarray}}
\newcommand{\req}[1]{(\ref{#1})}

\def\eps{\epsilon}

\begin{document}

\twocolumn[\hsize\textwidth\columnwidth\hsize\csname
@twocolumnfalse\endcsname

\title{Optimal combinations of imperfect objects}
\author{Damien Challet$^1$ and Neil F. Johnson$^2$}
\address{$^1$ Theoretical Physics, $^2$ Clarendon Laboratory, Physics
Department, Oxford University, Oxford OX1 3NP, U.K.}

\date{\today}
\maketitle

\begin{abstract} We consider how to make best use of
imperfect objects, such as defective analog and digital components. We show
that perfect, or near-perfect,  devices can be constructed by taking
combinations of such defects. Any remaining objects can be
recycled efficiently. In addition
to its practical applications, our `defect combination problem'
provides a novel generalization of classical optimization problems.
\vskip0.2in
\end{abstract} ]

Imperfection is an integral part of Nature, but it cannot always be
tolerated. High-technology devices, for example, must be precise and dependable. The
design of such dependable devices is the domain of fault-tolerant computing, where the
goal is to optimize reliability, availability or efficiency of redundant systems
\cite{BWJohnson}. Such redundant systems are typically built
from  devices which are initially defect-free and hence pass the quality
check, but may later develop faults. 

A much less studied problem, but one of significant economic and
ecologic importance, is what to do with a component which is already known to be
defective. Components with minor defects are sometimes acceptable for low-end
devices \cite{IEEE}. The Teramac, a massively parallel computer, was
built from partially defective conventional components; it uses an
adaptive wiring scheme between the components in order to avoid the defects,
and the
wires themselves can be defective \cite{teramac}.
More typically, however, a component that is known to be defective is considered
`useless' and is hence wasted.

Here we
address this wastage issue by relating it to a
novel optimization problem: given
a set of $N$ imperfect components, find a {\em combination}, or subset, that
optimizes the average error (analog components), or the number of
working transformations (digital components). We employ methods from statistical
physics to show that perfect, or near-perfect, devices can indeed be constructed, and
remaining objects can be recycled efficiently with (almost) zero net wastage. Note however that
combining {\em simple} analog devices such as thermometers, is not attractive since it
is usually much easier and cheaper to subtract the errors from
the outputs. But such active error-correction may not be practical in more complex
systems, particularly for next-generation technologies in the
ultrasmall nano/micro regime.  Nanoscale devices such as Coulomb-blockade
transistors may enable us to push back
the limits of Moore's law (see Ref. \cite{sciam} for a review). However, the accuracy
of the current produced at a given analog voltage will depend sensitively on the
reliability of the nanostructure's fabrication.
Similarly, the discrete optical transitions in semiconductor quantum dots \cite{neil}
can provide useful digital components for nanoscale classical computing \cite{neil2}. 
However, digital switching can only occur if the
energy levels coincide with the external light frequency. The accuracy of these energy
levels also depends on the precision of fabrication. However even in
self-assembled quantum dot structures, such as the ground-breaking virus-controlled
self-assembly scheme of  Ref.
\cite{virus} where quantum dots are mass-produced, no two individual dots will ever be
identical - each will contain an inherent, time-independent systematic defect as
compared to the intended design. Yet it would be highly undesirable to discard such
nanostructures given the potential applications of such `bio-nano' structures.

Consider an analog device such as a nanoscale transistor,
registering a current
$A+a$ given a particular applied voltage, with  $A$ being the actual value and  $a$
being the systematic error\cite{multifunct}. Suppose fabrication has
produced a batch of
$N$ imprecise devices whose errors $a_i$ ($i = 1,2,\cdots N$) were created
when the objects were built and remain {\em constant}; this amounts to
drawing them from a known distribution $P(a)$. For simplicity, we suppose that $P(a)$
is  Gaussian with average  $\mu$ and variance
$\sigma=1$\cite{sigmanon1}. The most precise component has an error of order
$\frac{1}{N}$. What should one do with the others?
Generally speaking, one could combine them such that their defects compensate.
Computing the average of all components leads to an error of
order $\mu\pm\frac{1}{\sqrt{N}}$ which vanishes very slowly for large
$N$, and even then only if $\mu=0$. Nevertheless
this method has been used in many contexts throughout history, for example
by sailors who often took several clocks on board ship \cite{JJL}.

The optimal combination is actually obtained by taking a
well-chosen subset of the  $N$ components, i.e. a subset containing  $M\leq
N$ devices whose errors
compensate best.  The problem
therefore consists of selecting some of the numbers $a_i$  such that the
absolute value of their average is minimized: hence the interesting
quantity is
\be\label{DCP}
\eps=\min_{\{n_i\}}\frac{|\sum_{j=1}^N n_j a_j|}{\sum_{k=1}^N
n_k}=\min_{\{n_i\}}\frac{E}{M}\ \ .
\ee
Here  $n_j\in\{0,1\}$  selects whether device $j$ is used or
not, while $M=\sum_{i=1}^N n_i$  is the total number of devices used. 
Without division by $M$,  this problem --- which we call the defect
combination problem (DCP) --- would be similar to the subset sum
problem or number
partitioning problem (NPP) \cite{GJ}.  Both problems
are equivalent and known to be
NP-complete: in the worst case, there is no method which finds the minimum
in polynomial
time, i.e. significantly faster than brute force enumeration (exponential
time). 
However the typical, i.e. average, problem has a different behavior. It will
undergo a transition between
a computationally hard phase where the average error is greater than zero,
and a computationally
easy phase where the error is zero \cite{KKLO,MertensPRL}. The same applies in our
present case. These two phases,
and the transition between them, can be studied using
statistical physics \cite{KKLO,MertensPRL,MertensPRL2,Z}.

\begin{figure}
\centerline{\psfig{file=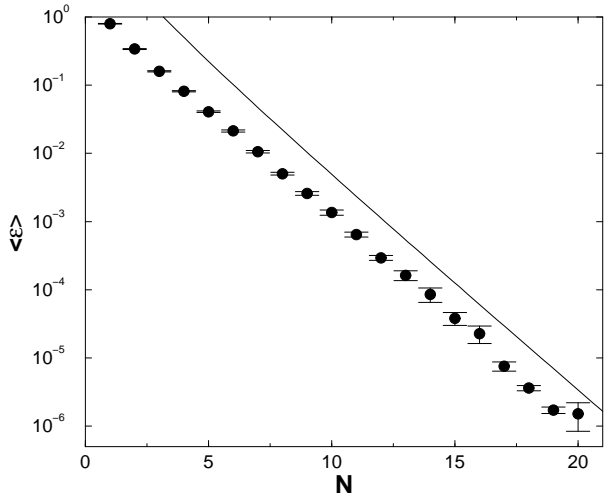,width=8cm,angle=0}}
\caption{Unconstrained problem: the average error   $\avg{\eps}$
versus the total number of components  $N$ obtained by
numerical evaluation. Average taken over 20000 samples. The solid line shows
the behavior of the theoretical upper bound, for a particular choice
of the constant $C$. } \label{unconstr}
\end{figure}

Figure \ref{unconstr} reports numerical results
obtained by enumerating all possible $\{n_i\}$, where 
$\mu=0$. The resulting device precision can be quite remarkable.
Our numerical simulations also confirm that
$\avg{M}=N/2$ for large $N$,  i.e. the optimal
configuration uses half
the components on average. 
Strong fluctuations remain even for a large number of
realizations and for large $N$, because of the non
self-averaging nature of the problem:
i.e. $\sqrt{\avg{\eps^2}-\avg{\eps}^2}/\avg{\eps}\to 1$.

The division by $M$ in Eq \req{DCP} makes the DCP
much harder to tackle analytically than the NPP. Let us
compare the DCP with ${\rm min}E$, the problem defined by finding the minimum of
$E$. Numerical simulations show that for the latter problem
$\avg{M^*}_{a,\text{minE}}=\frac{N-1}{2}$ for sufficiently large $N$,
where $M^*=\sum_{j=1}^N n_j^*$ is the number of selected components in
the configuration $\{n_i^*\}$ that minimizes $E$ \cite{argmin}. 
This
makes sense, since in the DCP the division by $M$ favors
configurations with a larger number of components. In addition
both problems are related by the inequality
\be
\min_{\{n_i\}}\cro{\frac{E}{M}}\le\frac{\min_{\{n_i\}}E}{M^*}\ \ .
\ee
Hence
\be
\avg{\eps}\le \frac{C}{N}\,\avg{\min_{\{n_i\}}E}
\ee
for some constant $C$. Computing the typical properties 
of ${\rm min}E$ hence yields an upper bound
to the average optimal error. Following Ref.
\cite{MertensPRL}, we computed the partition function
$Z=\sum_{\{n_i\}}\e^{-\beta|\sum_j n_j a_j|}$, which for large $N$ yields
\be
Z\simeq 2^N\int_{\pi/2}^{\pi/2}\e^{NG(y)}
\ee
where $G(y)=\avg{\ln(\cos(a\beta \tan(y)/2))}$. Using the saddle point
approximation for $Z$ and an argument of positive entropy \cite{MertensPRL},
we find
 $\avg{\min_{\{n_i\}}E}\simeq
\sqrt{\frac{\pi N}{2}} \,2^{-N}$. Hence there is a constant $C$ such that
\be
\avg{\eps}\le  C\,\frac{1}{\sqrt{N}\,2^{N}}.
\ee
Figure \ref{unconstr}  shows the behavior of the
analytically obtained upper-bound for the average error is consistent with
the corresponding numerical results. The same calculus shows that
the DCP will also  exhibit a phase transition
between hard and easy problems when $b<N-\ln(\pi N/6)/2$, where $b$ is 
the number of bits
needed to encode the $a_i$'s. Hence it
is possible to obtain perfect error-free combinations of such
imperfect objects for  $N$ large and  $b$ sufficiently small.
When the defects are biased, i.e.  $\mu\ne0$, the error increases as
$\mu$ increases but remains low for $\mu\le 1$. When the
errors of the components all have the same sign, only one component is used
and the
resulting error increases linearly with  $\mu$.

\begin{figure}
\centerline{\psfig{file=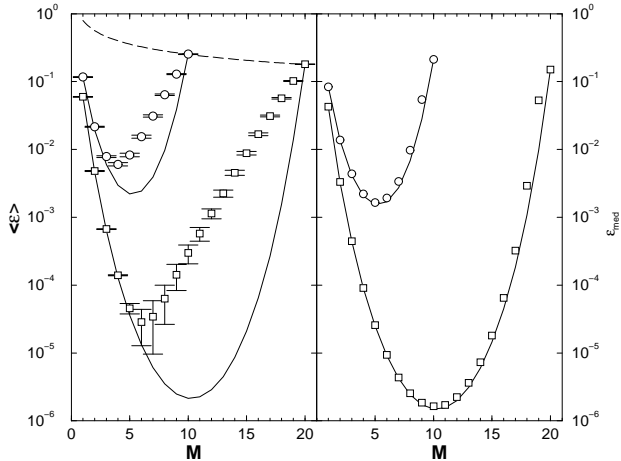,width=8cm,angle=0}}
\caption{Constrained problem. Left panel: average optimal error
$\avg{\eps}$  
versus the number of components  $M$  for a total of  $N = 10$
(circles) and 20 (squares) (average taken over 20000 samples). Right
panel: median error versus $M$. Continuous lines are the
predictions from the random cost approach. The dashed line is
$\sqrt{\frac{2}{\pi N}}$, the average error is taken over all
components.}
\label{constr}
\end{figure}

We now consider the constrained DCP, where the number of components to be used is
pre-defined to be 
a particular value $M$. If $M = 1$, one
selects the least imprecise component. The case  $M=N$  amounts to computing
the average over all
$N$ components, hence $\avg{\eps}=\sqrt{2/(\pi N)}$. 
This problem is a more complicated version of the subset sum problem:
in our case the numbers $a_i$ are no longer restricted to be
positive and the cost function is the absolute value of the
sum. Fig. \ref{constr} plots the average and median optimal error
as a function of $M$ for $N=10$ and 20. An exponential fit of the
minimum error for increasing $N$ gives $\min_M\avg{\eps}\sim
\exp(-K N)$ where $K=0.56\pm0.01$, to be compared with $-\ln 2=0.693\ldots$ in the
unconstrained case; the functional form of this
quantity may however be more complicated than an exponential.

We have also applied Derrida's random cost
approach \cite{Derrida,rancost,MertensPRL2}. Given the $N$ errors
$\{a_i\}$, there corresponds one $E$ to each of the $L={N \choose M}$ sets $\{n_i'\}$
that obey the constraint
$\sum_i n_i'=M$ \cite{Derrida,rancost,MertensPRL2}. 
If the $E$ are independent, all properties of the problem are then given by
the p.d.f. $p_M(E)$.
In our case, the latter is straightforward to compute:
\be
p_M(E)={N\choose M}^{-1}\sum_{\{n_i'\}}\avg{\delta(E-|\sum_jn_ja_j|)},
\ee
where the prime means that  $\sum_i n_i'=M$. Hence $p_M(E)$ is equal to the
probability
distribution of the absolute
value of $N$ numbers drawn from $P(a)$, which is 
\be
p_M(E)=\frac{\Theta(E)}{\sqrt{2\pi M}}\,\cro{\e^{-\frac{(E-\mu
M)^2}{2M}}+\e^{-\frac{(E+\mu M)^2}{2M}}}\ \ .
\ee
Let us concentrate on the non-biased case $\mu=0$ (the calculus is
easily extended to the biased case). We are interested in the average
value of the minimum $E_1$ of ${N\choose M}$ numbers drawn
from $P_M(E)$. Using
$P(E_1)=-\frac{\dd}{\dd E_1}[F_>(E_1)]^{L}$,
where $F_>(E_1)=\erfc(E_1/\sqrt{2M})$
is the cumulative distribution function of $P_M(E)$, we find that
\be\label{H_1rancost}
\frac{\avg{E_1}}{M}=\sqrt{\frac{2}{M}}\,\int_0^\infty \dd t
[\erfc(t)]^{N\choose M}\ \ .
\ee
If $M=N$, one recovers the average over all components since  $\int_0^\infty \dd t\, \erfc(t)=1/\sqrt{\pi}$. By
definition, the median is given by $\eps_{\text{med}}$ such that
$\int_0^{\eps_{\text{med}}}\dd E_1\,P(E_1)=1/2$.

The left panel of Fig. \ref{constr} plots the average constrained error,
obtained by numerical enumeration, and the analytical predictions of Eq
\req{H_1rancost}, for two sizes of component set.
The larger the component pool  $N$, the
better the precision.  For $M<<N$
the random cost approach describes
$\avg{\eps}$ well, however it fails dramatically for larger values of
$M$. This is because the $L={N\choose M}$ values of $E$
become increasingly dependent as $M/N$ grows. At fixed $N$, as $M$ increases, 
particular samples have an increasing probability to contain a large
fraction of defects  $a_i$ with  the same sign. Due to
the constraint on $M$, one may therefore be forced to use components  whose defects
add instead of compensating each other. The
median, which is less affected by such events,  has its mimimum close to
$N/2$ (right panel of Fig. \ref{constr}): the random cost approach describes much 
better the
behavior of the median than that of the average, although the discrepancy
increases as $M/N$ increases.

\begin{figure}
\centerline{\psfig{file=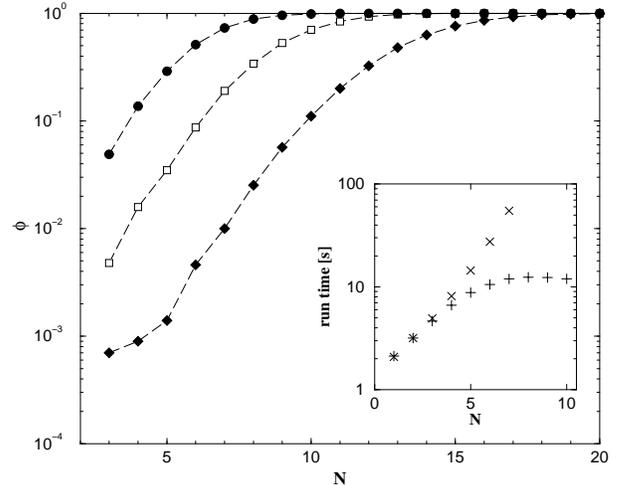,width=8cm,angle=0}}
\caption{Binary components:  average fraction $\avg{\phi}$ of perfect
devices as a
function of $N$ for $P=100$ and $f=0.2$ (filled circles), $f=0.25$ (squares)
and
$f=0.3$ (filled diamonds). Inset: running time versus $N$ of simple
enumeration
(x) and enumeration with sorting ($+$) ($f=0.2$). Averages taken over
$10000$ samples. Dashed
lines are for guidance to the eye.}
\label{bin}
\end{figure}

Using a subset of defective components is also a
powerful method for binary components such as quantum-dot optical switches
\cite{sciam,neil2}. Suppose each component has
$I$ input bits. If it can perform  $F$ different logical operations on the input
bits, it can perform  $P=F2^I$ different transformations (i.e. truth table
has
$P$ entries). Let $f$ be the
probability that for a given transformation $l$, component $i$ systematically gives 
the wrong output. e.g. because of inaccuracies in the energy-level spacings in the case
of quantum dot switches. Mathematically,
$P(a_i^l=-1)=f$,
$-1$ labelling a defective ouput and $1$ a correct one.
It becomes exponentially unlikely that one can extract a
perfect component as $f$ increases, however subsets of the components may
indeed produce the correct output. One therefore selects a subset of
components from a pool of
$N$, in order to maximize the number of transformation such that the
majority of components give
the correct answer. The maximal fraction of working
transformations for a given set of components $\{a_i\}$ is
\be
W(\{a_i\})=\frac{1}{P}\,\max_{\{n_i\}}\,\sum_{l=1}^P\Theta(\sum_{j=1}^Na_j^l
n_j)
\ee
where $\Theta(x)$ is the Heaviside function. We measured
$\phi=\avg{\delta_{W,1}}$, the average fraction of component sets with at 
least one
perfect subset. Numerical calculations confirm that it is
indeed possible to build perfectly working components even if  $f$  is so high that no
single component is perfect.  Figure \ref{bin} shows numerical simulations of
the probability $\phi$ versus $N$ for three values of
$f$. When $\phi>0$, an efficient algorithm consists in first ranking (Heapsort) the
components according to the number
of working transformations and then enumerate all possibilities until a
perfect combination is found, beginning with the
less defective ones (see inset of Fig \ref{bin}). Analytic results along the lines of
 \cite{Derrida} will presented elsewhere.

Admittedly, seeking
optimal combinations implies an additional cost for two reasons.
First, one has to find the optimal or near optimal combination; this can be
done either
by measuring all the defects and then finding the minimal error with
a computer; or, skipping the labour-intensive step of measuring individual
defects,
by building combinations of objects such that we eventually minimize the
aggregate error. Second,
these objects have to
be wired, and their output combined by an additional hopefully error-free
device. 
However such wiring and selection of working subsets of components, is
precisely what is already being done
inside the Teramac \cite{teramac}. Given that defective components
can be cheaply produced en masse, the cost of such wiring and selection
of working combinations may not be an obstacle. Hence we believe
that our two optimization problems may prove relevant in practice, in
particular in emerging technologies
where the fabrication of defect-free components may not be possible.

Our scheme implies that the `quality' of a component is not determined
solely by its own intrinsic error.
Instead error becomes a collective property, which is determined by the
`environment' corresponding to the
other defective components. Efficient recycling of otherwise `useless'
components now becomes possible.
Suppose that a fabrication process produces a
constant flow of defective analog or binary components. One can now perform
the
following scheme to generate a continuous output of useful devices: fix
$N$  according to the desired average error (see Fig. \ref{unconstr}, Fig.
\ref{constr} or Fig \ref{bin}); form the optimal subset; add fresh
components to the
unused ones; find the optimal subset, and repeat as desired. The
quality of the subset fluctuates, but there is essentially no wastage.
Although efficient
algorithms for the analog case remain to be found,
generalization of well-known algorithms
\cite{KK} may be
possible. We hope that our work inspires further academic research into this
important practical problem.

We thank R. Zecchina and D. Sherrington
for useful comments. D.C. thanks the Swiss National Funds for
Scientific Research
for financial support.


\begin{thebibliography}{99}
\bibitem{multifunct} Our approach is easily extended to multifunction
devices.
\bibitem{sigmanon1} All results can be extended to the general case
$\sigma<\infty$ simply by multiplying $\eps$ and $H$ by $\sigma$.
\end{thebibliography}
\end{document}